\documentclass[twocolumn]{autart}    

\usepackage{cite,color}
\usepackage{amsmath,amssymb,amsfonts}
\usepackage[bb=boondox]{mathalfa}
\usepackage{comment}
\usepackage{graphicx}
\usepackage{subcaption}
\usepackage{setspace}

\usepackage{tikz}
\usetikzlibrary{arrows.meta, positioning}
\usetikzlibrary{decorations.pathmorphing, decorations.markings, positioning, calc, backgrounds}
\usepackage{pgfplots}
\pgfplotsset{compat=1.18}

\newtheorem{secthm}{Theorem}[section]
\newtheorem{seccor}[secthm]{Corollary}

\newtheorem{secex}[secthm]{Example}
\newtheorem{secprop}[secthm]{Proposition}
\newtheorem{secdefn}[secthm]{Definition}

\newtheorem{secasm}[secthm]{Assumption}
\newtheorem{secprob}[secthm]{Problem}

\newcommand{\He}[1] {\mbox{\rm He}\left(#1\right)}
\newcommand{\bR} { {\mathbb R}}
\newcommand{\bS} { {\mathbb S}}

\newcommand{\0}{\mathbb{0}}
\newcommand{\I}{\mathbb{I}}

\renewcommand{\theenumi}{\roman{enumi}}

\def\red{\hfill $\lrcorner$}

\definecolor{c1}{RGB}{31,119,180}
\definecolor{c2}{RGB}{255,127,14}
\definecolor{c2}{RGB}{255,187,120}
\definecolor{c3}{RGB}{44,160,44}
\definecolor{c4}{RGB}{214,39,40}
\definecolor{c5}{RGB}{148,103,189}
\definecolor{c6}{RGB}{140,86,75}
\definecolor{c7}{RGB}{227,119,194}
\definecolor{c8}{RGB}{127,127,127}
\definecolor{c9}{RGB}{188,189,34}
\definecolor{c10}{RGB}{23,190,207}

\allowdisplaybreaks[4]

\begin{document}

\begin{frontmatter}

\title{
Design of MIMO Lur'e oscillators via dominant system theory \\
with application in multi-agent rhythm synchronization\thanksref{footnoteinfo}
}

\thanks[footnoteinfo]{This paper was not presented at any IFAC meeting. 
The work of Y. Kawano was partially supported by JST FOREST Program Grant Number JPMJFR222E and JSPS KAKENHI Grant Number JP24K00910. 
Corresponding author Y.~Kawano. Tel. +81-82-424-7582. Fax +81-82-424-7193.}

\author[JP]{Yu Kawano}\ead{ykawano@hiroshima-u.ac.ip},  
\author[UK]{Fulvio Forni}\ead{f.forni@eng.cam.ac.uk},  

\address[JP]{Graduate School of Advanced Science and Engineering, Hiroshima University, Higashi-Hiroshima 739-8527, Japan} 

\address[UK]{Department of Engineering, University of Cambridge, Cambridge, CB2 1PZ, UK} 



\begin{abstract}
This paper presents a new design framework for dynamic output-feedback controllers for Lur’e oscillation in a multiple-input multiple-output setting. We first revisit and extend dominant system theory to state-dependent rates, with the goal of deriving conditions based on linear matrix inequalities. Then, we introduce a separation principle for Lur’e oscillator design, which allows for the independent design of a state-feedback oscillator and an observer. Our proposed control synthesis is demonstrated through the rhythm synchronization in multi-agent systems, illustrating how networks of stable, heterogeneous linear agents can be driven into phase-locked rhythmic behavior.
\end{abstract}

\end{frontmatter}





\section{Introduction}

Oscillators are at the core of many modern electro-mechanical devices, from simple integrated circuits to complex systems for power conversion and robot locomotion~\cite{KAS:99,Ijspeert:08}. Feedback plays a crucial role in all oscillator mechanisms. The design of harmonic oscillators, for instance, requires a loop gain of at least one and a total phase shift of at least 180 degrees to guarantee sustained oscillations. In engineering, harmonic oscillator design is supported by well-established methods, such as the describing function method~\cite{KB:50,MB:75,GV:68,Iwasaki:08}. At the other end of the spectrum, relaxation oscillators operate by a slow buildup of energy in a storage element. This energy eventually reaches a threshold, triggering an internal switch that causes rapid discharge. For 
a system-theoretic analysis of relaxation oscillators, we refer to~\cite{Astrom:95,VG:02,GMD:01}.

Although the design of oscillators is a well-established art in electronics \cite{Gottlieb:97,Gonzalez:06}, in system theory the available methods pale in comparison to the more classical problem of equilibrium stabilization. The difference with equilibrium stabilization is that the existing methods for characterizing oscillations, such as Poincar\'e-Bendixson theory \cite{HDS:74} and Hopf bifurcations \cite{AVK:66}, do not offer similar tractability or scalability. In fact, an algorithmic, optimization-based design for oscillators is currently lacking. This paper aims to present some results in this direction. We restrict our study to the combination of linear dynamics and simple static nonlinearities, as in the Lur'e setting, to derive linear matrix inequalities (LMIs) for the automated synthesis of oscillators.

\begin{figure}[b!]
    \centering
    \includegraphics[width=0.9\linewidth]{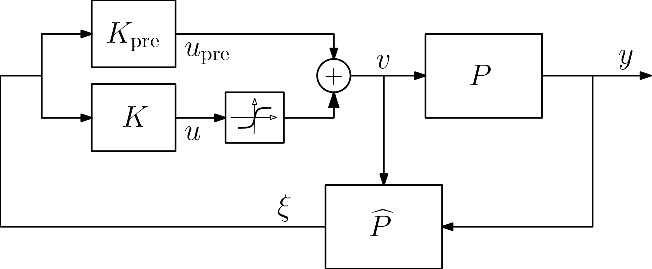}
    \caption{Closed-loop block diagram for controlled oscillations.}
    \label{fig:block}
\end{figure}

We focus on the closed-loop system shown in Fig. \ref{fig:block}. Here, $P$ represents a stable linear plant, and  
$\hat{P}$ is a state-estimator. Our goal is to design the state-feedback gains, $K_{\mathrm{pre}}$ and $K$, to ensure oscillations. The output of one of the two controllers is filtered by a sigmoidal static nonlinearity. Saturations are common nonlinear behaviors and a natural choice for many oscillator studies, especially for Lur'e oscillators. In fact, by extracting the nonlinearity and combining the linear elements, the diagram of Fig. \ref{fig:block} can be analyzed within the Lur'e setting.

Several additional reasons motivated the choice of this architecture. First, the inclusion of an observer is driven by our goal of deriving a dynamic output feedback design, a topic seldom explored in the nonlinear control of oscillators. This approach also allows us to extend the separation principle of linear stability theory to oscillators. This would guarantee that oscillations can be established by a combination of a state-feedback controller and a dynamic observer, both designed independently.
The parallel branches of the state feedback, identified by signals $u$ and $u_{\mathrm{pre}}$, are intended to combine linear and nonlinear actions in a way that retains control authority across all operating regimes. The control action $u$ is limited in range, and thus has authority only in the small-signal regime. The large-signal regime is dominated by $u_{\mathrm{pre}}$.
Finally, when considering the complete controller (comprising observer, parallel static branches, and the sigmoidal nonlinearity), we can see that the nonlinear elements are minimal. All the tunable elements of the controller are linear. This design choice is motivated by tractability, with the goal of deriving computable conditions for the design of oscillators.

This paper leverages the recent dominant system theory \cite{MFS:18,FS:19} which we extend to include state-dependent rates, broadening its applicability. We derive new conditions for oscillator design for multiple-input multiple-output (MIMO) plants. Through relaxation, these conditions lead to linear matrix inequalities (LMIs) for oscillator design. We also propose a separation principle to combine state feedback with state estimation, beyond equilibrium stability. This allows us to derive both state-feedback and dynamic output-feedback controllers for oscillators. 
We provide several examples of stable systems controlled into self-sustained oscillations via feedback. Moreover, we illustrate the theory's potential on multi-agent systems, showing how networks of stable, heterogeneous linear agents can be driven into phase-locked rhythmic behavior. The heterogeneous dynamics mean that each agent follows a different trajectory, yet they agree on a common rhythm and phase-lock. Notably, these oscillations disappear as soon as the controller is disconnected, which is an important feature of our design.

The results of this paper show contact points with the results in recent papers \cite{SKW:23,CF:24,JF:25}, which focus on state-space methods for oscillators design and make use of dominant system theory.
However, for the first time this paper utilizes the problem structure of the general MIMO setting of Fig. \ref{fig:block} and introduces and leverages extended dominant system theory, 
to derive new technical conditions for oscillations. 

The paper is organized as follows: Section \ref{sec:DLO} formalizes the design problem. Section \ref{sec:dominant_nutshell} recalls and extends the theory of dominant systems. State-feedback and dynamic output-feedback solutions for oscillations are detailed in Sections \ref{sec:state_feedback} and \ref{sec:output_feedback}, respectively. Each of these sections provides the main theoretical conditions and then discusses various implementation methods using convex optimization. Section \ref{sec:multiagent} focuses on the design of phase-locked multi-agent oscillator networks. Nonlinear extensions to the block diagram in Fig. \ref{fig:block} are discussed in Section \ref{sec:nonlinear}, with particular emphasis on open challenges. Conclusions follow. All proofs are in the appendix.

{\it Notation:}
The sets of real numbers, non-negative real numbers, and positive real numbers are denoted by $\bR$, $\bR_{\ge 0}$, and $\bR_{> 0}$, respectively.
The set of $n \times n$ symmetric matrices is denoted by $\bS_n$.
The $n$-dimensional vector and $n \times m$-matrix whose all components are $0$ are denoted by $\0_n$ and $\0_{n \times m}$, respectively.
The $n \times n$ identity matrix is denoted by $\I_n$.
For $A \in \bR^{n \times n}$, $\He{A} := A + A^\top$.
For $x \in \bR^n$, ${\rm diag}(x)$ denotes the diagonal matrix whose $i$th diagonal element is the $i$th element of $x$.
For $P, Q\in \bS_n$, $P \succ Q$ (resp. $P \succeq Q$) means that $P - Q$ is positive (resp. semi) definite. A matrix $P \in \bS_n$ is said to have inertia $p$ if it has $p$ negative real eigenvalues and $n-p$ positive real eigenvalues.
For $f:\bR^n \to \bR^n$ of class $C^1$, $\partial f(x)$ denotes the Jacobian matrix of $f(x)$ computed at $x \in \bR^n$



\section{Problem Formulation}\label{sec:DLO}
Consider the feedback loop represented in Fig. \ref{fig:block}
given by a linear time-invariant (LTI) plant
\begin{subequations}
\label{eq:sys}
\begin{align}
\label{eq:linear}
\left\{
\begin{alignedat}{2}
\dot x &= A x + B v \\
y &= C x
\end{alignedat}
\right.
\qquad
x \in \bR^n, 
v \in \bR^m,
y \in \bR^q,
\end{align}
whose control input $v$ is given by
\begin{align}
\label{eq:static_v}
v = u_{\mathrm{pre}} + \varphi(u) ,
\qquad u_{\mathrm{pre}} \in \bR^m, u \in \bR^m, 
\varphi: \bR^m\to\bR^m
\end{align}
Both control signals $u_{\mathrm{pre}}$ and $u$ are
generated by linear controllers, given by  the combination of
static gains
\begin{align}
\label{eq:static_u}
u_{\mathrm{pre}} = K_{\mathrm{pre}} \xi,
\qquad u = K \xi,
\qquad K_{\mathrm{pre}},K \in \bR^{m\times n},
\end{align}
and observer dynamics
\begin{align}
\label{eq:obs_dyn}
\dot \xi = A \xi + B v + L(C\xi-y), \qquad
\xi \in \bR^n, L \in \bR^{n\times q}.
\end{align}
\end{subequations}
Signal $u$ is filtered by 
a sigmoidal nonlinearity $\varphi$,
which satisfies the following assumption.
 \begin{secasm}
 \label{secasm:static_nonlinearity}
 $\varphi : \bR^m\to\bR^m$
 is diagonal as in
 \begin{subequations}
 \begin{equation}\label{eq:diagonal}
 \varphi (u) := \begin{bmatrix} \varphi_1 (u_1) & \cdots & \varphi_m (u_m) \end{bmatrix}^\top.
 \end{equation} 
 For each $i = 1, 2, \dots, m$,
 $\varphi_i: \bR \to \bR$ is a class $C^1$ bounded function such that
\begin{align}
\varphi_i (0) = 0,
\end{align}
and there exists $\overline{\theta}_i \in \bR_{>0}$ such that
\begin{align}\label{eq:bound}
\partial \varphi_i (u_i) \in [0, \overline{\theta}_i]
\end{align} 
\end{subequations}
for all $u_i \in \bR$. 
\red
\end{secasm} 

Our goal is to design a feedback controller that guarantees stable closed-loop oscillations. We first solve the simpler \textbf{static feedback} case, where $y=x$ and \eqref{eq:static_u} and \eqref{eq:obs_dyn} are replaced by
\begin{align}
\label{eq:static_u_state_feedback}
u_{\mathrm{pre}} = K_{\mathrm{pre}} x, \qquad u = K x.
\end{align}
 We then consider the \textbf{output feedback} case
\eqref{eq:linear}-\eqref{eq:obs_dyn}.

\begin{secprob}\label{prob}
Consider the plant \eqref{eq:linear}
driven by the input \eqref{eq:static_v}.
Under Assumption \ref{secasm:static_nonlinearity},
find either
\begin{enumerate}
\item[(i)] the static feedback controller 
\eqref{eq:static_u_state_feedback}
or 
\item[(ii)]  the dynamic feedback controller
\eqref{eq:static_u},\eqref{eq:obs_dyn}
\end{enumerate}
that guarantees that almost all trajectories of the closed-loop system converges to some limit cycle.~\red
\end{secprob}

\section{Extended Dominant System Theory}
\label{sec:dominant_nutshell}

\begin{secdefn}
Consider the closed nonlinear system
\begin{align}\label{eq:asys}
\dot x = f(x),
\end{align}
where $f: \bR^n \to \bR^n$ is of class $C^1$.

Take $\lambda : \bR^n \to \bR_{\geq 0}$.
The system~\eqref{eq:asys} is said to be \emph{strictly $p$-dominant} with rate $\lambda (x)$ 
if there exist $\varepsilon  > 0$ and $X \in \bS_n$ with inertia $p$ such that
\begin{align}\label{eq:domi}
\He{X (\partial f(x) + \lambda (x) \I_n)} + \varepsilon  \I_n \preceq \0_{n \times n}
\end{align}
holds for all $ x \in \bR^n$.
\red
\end{secdefn}

This definition extends \cite[Definition 2]{FS:19} to non-constant rates~$\lambda (x)$.
The inequality~\eqref{eq:domi} implies that $n-p$ eigenvalues of $\partial f(x) + \lambda (x) \I_n$ lie in the left-half plane while the other $p$ eigenvalues do in the right-half plain. The intuition is that the system behavior is determined by an asymptotically stable system of dimension $n-p$ and a richer, not necessarily converging dynamics of dimension $p$. This intuition is formalized in the next
proposition, which extends \cite[Corollary 1]{FS:19}.
The original statement is for constant rates~$\lambda$.

\begin{secprop}\label{prop:domi}
Let \eqref{eq:asys} be a strictly $p$-dominant system with rate 
$\lambda : \bR^n \to \bR_{\geq 0}$.
Then, any bounded trajectory of \eqref{eq:asys}
converges to
\begin{itemize}
\item a unique fixed point if $p=0$;
\item a fixed point if $p=1$;
\item a simple attractor (i.e., a fixed point, a set of fixed points, or a periodic orbit) if $p=2$.
\red
\end{itemize}
\end{secprop}
\begin{pf}
The proof of \cite[Corollary 1]{FS:19} can easily be extended
to Proposition \ref{prop:domi} by taking
$\lambda : \bR^n \to \bR_{\geq 0}$.
\qed
\end{pf}

We take advantage of Proposition~\ref{prop:domi} to solve Problem~\ref{prob}. 
The idea is to design controllers that guarantee the closed-loop system
\begin{itemize}
\item has bounded trajectories,
\item is strictly $2$-dominant for some rate $\lambda (x) \ge 0$,
\item has only one equilibrium and is unstable.
\end{itemize}
In agreement with Proposition 3.2, under these conditions,
every closed-loop trajectory must converge to a simple attractor. Since there is only one unstable equilibrium, almost all trajectories must converge to some limit cycle.

\begin{figure}[h]
\centering
\includegraphics[width=.9\columnwidth]{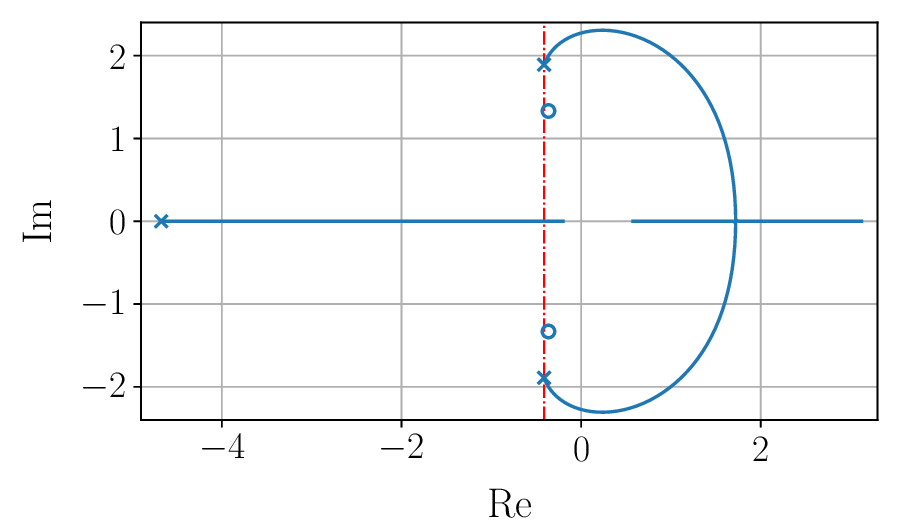} 
\caption{Root locus of \eqref{eq:rlocus} for $A$, $B$, and $K$ in \eqref{eq:ABK} for $\theta_1 \in [0, 1]$}
\label{fig:rlocus}
\end{figure}

\begin{secex}\label{ex:lam}
In this example, we illustrate the benefit of utilizing non-constant $\lambda (x)$.
Consider 
\begin{subequations}\label{eq:example:3.3}
\begin{align}
f(x) &= Ax + B \tanh(Kx) \\
\label{eq:ABK}
A &=
\begin{bmatrix}
-3 & -2.2 & -0.6\\
0 & 0.5 & 1\\
-4 & -4 & -3
\end{bmatrix},
\quad 
B = 
\begin{bmatrix}
3 \\ 0.2 \\ 6
\end{bmatrix}
\nonumber\\
K &= 
\begin{bmatrix}
1 & 0 & 1
\end{bmatrix}.
\end{align}
\end{subequations}
This system is strictly $2$-dominant with respect to 
\begin{align*}
\lambda (x) = 3(1 -  \partial \tanh (Kx))
\end{align*}
because \eqref{eq:domi} holds for $\varepsilon = 0.01$.
The eigenvalues of $X$ are $-3$, $-0.707$, and $1.26$, and thus $X$ has inertia $2$. 
Applying Theorem~\ref{thm:sfb} below, it is possible to show that almost all trajectories converges to some periodic orbit as in Fig.~\ref{fig:sim_ex3}.


\begin{figure}[htbp]
\centering
\includegraphics[width=0.9\columnwidth]{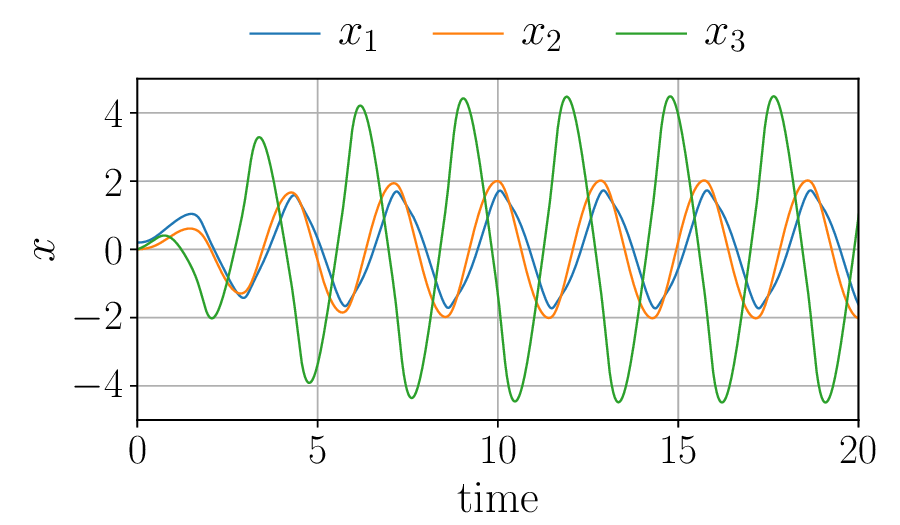} 
\caption{Solution of \eqref{eq:example:3.3}}
\label{fig:sim_ex3}
\end{figure}


Notably, for \emph{constant} $\lambda \ge 0$, solving~\eqref{eq:domi} with respect to $\varepsilon > 0$ and $X$ of inertia $2$ is necessarily infeasible. This can be explained by using the root locus of the feedback interconnection
\begin{subequations}\label{eq:rlocus}
\begin{align}
& \quad u = - \theta_1 y \\
&\left\{\begin{alignedat}{2}
\dot x & = A x + B u\\
y & = - K x.
\end{alignedat}\right.
\end{align}
\end{subequations}
The root locus describes the motion of the eigenvalues of $A + \theta_1 B K$ as $\theta_1$ goes from $\theta_1 = 0$ to $\theta_1 = \infty$. We observe that that the eigenvalues of the Jacobian matrix $\partial f(x) = A + \partial \tanh(Kx) B K$ correspond to the eigenvalues of $A + \theta_1 B K$ for $\theta_1 \in [0, 1]$. 
As shown in Fig.~\ref{fig:rlocus}, 
there is no vertical line that can split the eigenvalues  of $A + \theta_1 B K$ into two separated left and right groups, uniformly for $\theta_1 \in [0, 1]$. This is due to the overlap at $-0.412$ (the red dashed line). \
Therefore, the eigenvalues  of $A + \theta_1 B K + \lambda \I_3$ cannot be uniformly split into a group of unstable eigenvalues and a group of stable eigenvalues for any constant $\lambda \geq 0$, a necessary condition for $p$-dominance. This implies that \eqref{eq:domi} is infeasible for any constant $\lambda$. $\lambda (x)$ can address this issue.
\red
\end{secex}


\section{State Feedback Solution}
\label{sec:from2domLTI}
\label{sec:state_feedback}

\subsection{Theoretical Conditions for Oscillations}

This subsection is based on the key assumption 
that the pre-controller $K_{\mathrm{pre}}$ can make the system 
2-dominant for some given rate $\gamma>0$. This assumption guarantees that the inertia 
constraint of the linear matrix inequality (LMI) for closed-loop dominance presented below 
is automatically verified (for the rate $\lambda (x)$).

\begin{secasm}\label{asm:domi}
The matrix $A_{\mathrm{pre}} := A +B K_{\mathrm{pre}}$ is Hurwitz and
has exactly two eigenvalues whose real part is greater than $-\gamma$,
for some $\gamma > 0$.
\end{secasm}

To guarantee stable oscillations in closed-loop, 
we design $K$ to ``shift'' the two right-most eigenvalues of
$A_{\mathrm{pre}} + B \partial \varphi (Kx) K$ from stable to unstable while keeping the other eigenvalues negative real, as illustrated in Fig.~\ref{fig:rlocus}, where $\theta = \partial \varphi (Kx)$.

\begin{secthm}\label{thm:sfb}
Under Assumption~\ref{asm:domi}, let $\bar \varepsilon > 0$, $\bar K \in \bR^{m \times n}$, $\bar X \in \bS_n$, and $\bar \lambda:\bR^m \to [0, \gamma]$ with $\bar \lambda (\0_m) = 0$ be such that
\begin{subequations}\label{eq:domi_sfb}
\begin{align}
\He{(A_{\mathrm{pre}} + \gamma \I_n) \bar X} + \bar \varepsilon  \I_n &\preceq \0_{n \times n} \label{eq1:domi_sfb}\\
\He{(A_{\mathrm{pre}} \!+\! \bar \lambda (\bar u) \I_n) \bar X \!+\! B \partial \varphi (\bar u) \bar K} + \bar \varepsilon  \I_n &\preceq \0_{n \times n} \label{eq2:domi_sfb}
\end{align}
\end{subequations}
and
\begin{align}
{\rm rank}\left( A_{\mathrm{pre}} \bar X + B \partial \varphi (\bar u) \bar K \right) = n&
\label{eq:rank_sfb}
\end{align}
hold for all $\bar u \in \bR^m$.
Then, $K := \bar{K}\bar{X}^{-1}$ guarantees that almost all trajectories of the closed-loop system  \eqref{eq:linear},\eqref{eq:static_v},\eqref{eq:static_u_state_feedback} converge to some periodic orbit. 
\end{secthm}
\begin{pf}
The proof is in Appendix~\ref{app:thm:sfb}.
\qed
\end{pf}

The rank condition~\eqref{eq:rank_sfb} guarantees that none
of the eigenvalues of $A_{\mathrm{pre}} + B \partial \varphi (K x) K$ passes through the origin when in crossing the complex plane imaginary axis, as in Fig.~\ref{fig:rlocus}. 
Compared to the classical Hopf bifurcation~\cite{AVK:66}, which studies the existence of a stable periodic orbit
in a ``local'' neighborhood of the unstable equilibrium (and for eigenvalues that remain near the imaginary axis),
Theorem~\ref{thm:sfb} takes advantage of $2$ dominance to establish
the existence of generic closed-loop oscillations.

To design a state feedback for closed-loop oscillations based on Theorem~\ref{thm:sfb}, we need to handle an infinite family of LMIs~\eqref{eq:domi_sfb} with respect to $\bar u \in \bR^m$. This can be addressed through approximation, by taking a sufficiently large number of samples of $\bar u \in \bR^m$. Stronger formal guarantees can be obtained by deriving a finite family of LMIs via convex relaxation.

\begin{seccor}\label{cor:sfb}
Under Assumption~\ref{asm:domi}, let $\partial \varphi_i (0) = \overline{\theta}_i$, $i=1,\dots,m$. Let $\bar \varepsilon > 0$, $\bar K \in \bR^{m \times n}$, $\bar X \in \bS_n$, and $\hat \lambda:\{0, \overline{\theta}_1\} \times \cdots \times  \{0, \overline{\theta}_m\} \to [0, \gamma]$ with $\hat \lambda (\overline{\theta}_1,\dots,\overline{\theta}_m) = 0$ be such that
\begin{subequations}\label{eq:domi_sfb2}
\begin{align}
\He{(A_{\mathrm{pre}} + \gamma \I_n) \bar X} + \bar \varepsilon \I_n 
& \preceq \0_{n \times n}  \label{eq1:domi_sfb2}\\
\He{(A_{\mathrm{pre}} \!+\! \hat \lambda (\theta) \I_n ) \bar X \!+\! B {\rm diag}(\theta) \bar K} + \bar \varepsilon \I_n 
& \preceq \0_{n \times n}  \label{eq2:domi_sfb2}
\end{align}
\end{subequations}
hold for all $\theta \in \{0, \overline{\theta}_1\} \times \cdots \times  \{0, \overline{\theta}_m\}$, and
\begin{align}\label{eq:rank_sfb2}
{\rm rank}\left( A_{\mathrm{pre}} \bar X + B {\rm diag}(\theta) \bar K \right) = n
\end{align}
holds for all $\theta \in [0, \overline{\theta}_1] \times \cdots \times  [0, \overline{\theta}_m]$.
Then, $K := \bar{K}\bar{X}^{-1}$ guarantees that almost all trajectories of the closed-loop system \eqref{eq:linear},\eqref{eq:static_v},\eqref{eq:static_u_state_feedback} converge to some periodic orbit.
\end{seccor}
\begin{pf}
The proof is in Appendix~\ref{app:cor:sfb}.
\qed
\end{pf}

Since~\eqref{eq:domi_sfb2} is a finite family of LMIs, this is numerically tractable. In contrast, verifying~\eqref{eq:rank_sfb2} is still challenging. This aspect is elaborated in the next subsection. 

Corollary~\ref{cor:sfb} gives a weaker condition for strictly $2$-dominance than \cite[Theorem 7]{CF:24},
since we do not require $\hat \lambda (\theta) = \gamma$ for all $\theta \in \{0, \overline{\theta}_1\} \times \cdots \times  \{0, \overline{\theta}_m\}$ by virtue of utilizing non-constant $\lambda (x)$.
Also, we provide the condition~\eqref{eq:rank_sfb2} to guarantee closed-loop oscillation.



\subsection{Verification of Technical Conditions}\label{sec:verification}
Assumption~\ref{asm:domi} requires that exactly $2$ eigenvalues of the state matrix $A_{\mathrm{pre}}$ are in the vertical stripe of the complex plane between the two vertical asymptotes passing through $(-\gamma, 0)$,
while all the others are in $(-\infty, - \gamma)$. 
This can be satisfied for controlled systems if a pair $(A,B)$ is controllable, through the action of the \emph{pre-compensator} $K_{\mathrm{pre}}$,
as shown in Fig.~\ref{fig:block}. A simple pole-placement procedure can be used.

In Theorem~\ref{thm:sfb}, there is a flexibility for the selection of  $\gamma > 0$ when verifying~\eqref{eq:domi_sfb}. For sufficiently small $\bar \varepsilon > 0$, the LMI~\eqref{eq1:domi_sfb} has a solution $X$ if and only if $A_{\mathrm{pre}} + \gamma I_n$ have two positive real eigenvalues, and the others are negative real. 
Let ${\rm Re}(\lambda_i)$, $i=1,\dots,n$ denote the real part of each eigenvalue of $A_{\mathrm{pre}}$ with ordering ${\rm Re}(\lambda_i) \ge {\rm Re}(\lambda_{i+1})$. Assumption~\ref{asm:domi} implies $0 > {\rm Re}(\lambda_1)\ge {\rm Re}(\lambda_2) > {\rm Re}(\lambda_3)$. Thus, for any $\gamma \in (-{\rm Re}(\lambda_2), -{\rm Re}(\lambda_3))$, there exist $\bar \varepsilon > 0$ and $\bar X$ of inertia~$2$ such that \eqref{eq1:domi_sfb} holds. Also, there is flexibility in the selection of $\bar \lambda (\bar u)$ in \eqref{eq2:domi_sfb}. Although there is no constructive way of finding suitable $\gamma$ and $\bar \lambda (\bar u)$ at present, these flexibilities can be utilized when solving~\eqref{eq:domi_sfb}.

The rank condition \eqref{eq:rank_sfb2} is equivalent to that all eigenvalues of $A_{\mathrm{pre}}+ B {\rm diag}(\theta) K$ are non-zero for all $\theta \in [0, \overline{\theta}_1] \times \cdots \times [0, \overline{\theta}_m]$. This can be numerically verified by drawing the eigenvalues for all $\theta$. In the single-input (SI) case, the eigenvalues of $A_{\mathrm{pre}}+ \partial \varphi (Kx) B K$ can also be visualized by drawing the root locus of \eqref{eq:rlocus}, which gives the following corollary of Theorem~\ref{thm:sfb}.

\begin{seccor}\label{cor2:sfb}
Under Assumption~\ref{asm:domi}, consider $m=1$ and $\partial \varphi (0) = \overline{\theta}_1$. Let $\bar \varepsilon > 0$, $\bar K \in \bR^{1 \times n}$, and $\bar X \in \bS_n$ be such that 
\begin{subequations}\label{eq:domi_sfb2_SI}
\begin{align}
\He{(A_{\mathrm{pre}} + \gamma I_n) \bar X} + \bar \varepsilon \I_n &\preceq 0 \label{eq1:domi_sfb2_SI}\\
\He{A_{\mathrm{pre}} \bar X + B \bar K} + \bar \varepsilon  \I_n & \preceq 0 \label{eq2:domi_sfb2_SI}\\
KA_{\mathrm{pre}}^{-2}B & \neq 0.
\end{align}
\end{subequations}
Then, $K := \bar{K}\bar{X}^{-1}$ guarantees that almost all trajectories of the closed-loop system \eqref{eq:linear},\eqref{eq:static_v},\eqref{eq:static_u_state_feedback} converge to some periodic orbit.
\end{seccor}
\begin{pf}
The proof is in Appendix~\ref{app:cor2:sfb}.
\qed
\end{pf}

Corollary~\ref{cor2:sfb} implies that if the set of \eqref{eq1:domi_sfb2_SI},\eqref{eq2:domi_sfb2_SI} is feasible, then we can always achieve closed-loop oscillation in the SI case. We observe that if $K A_{\mathrm{pre}}^{-2} B =0$, a (sufficiently) small perturbation $K':=K + \alpha B^\top A_{\mathrm{pre}}^2$ for $\alpha \neq 0$ guarantees that $K'A_{\mathrm{pre}}^{-2}B \neq 0$. In addition, from the continuity of the closed-loop eigenvalues with respect to $\alpha$, for sufficiently small $|\alpha| \neq 0$, there exist $\varepsilon > 0$ and $X \in \bS_n$ with inertia $2$ such that the closed-loop system satisfies~\eqref{eq:domi} with the same rate $\lambda (x)$.

\begin{figure}[htbp]
\centering
\includegraphics[width=0.9\columnwidth]{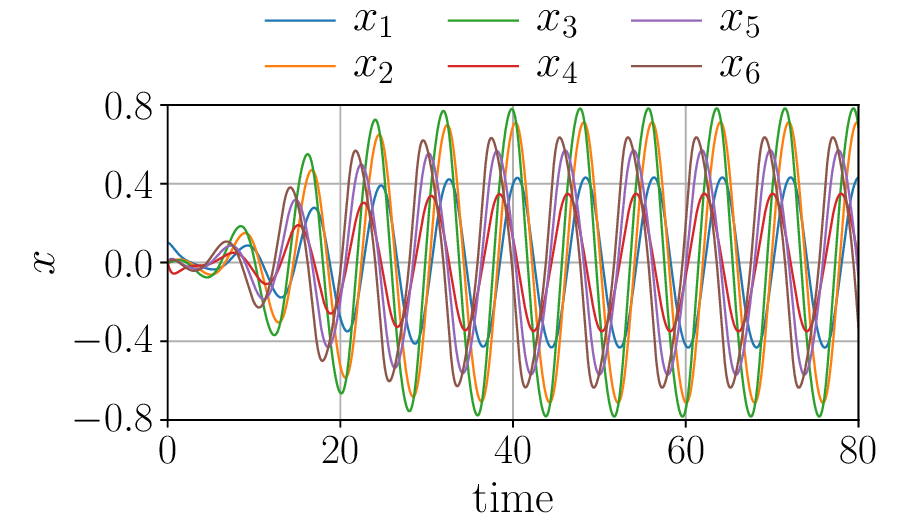}
\caption{Soluition of a simplified robotics model}
\label{fig:mech_state}
\end{figure}

\begin{secex}\label{ex:mech}
In this example, we illustrate Corollary~\ref{cor2:sfb}.
The mass-spring-damper system in Fig.~\ref{fig:mdk_linear} provide a simplified robotics model for peristaltic locomotion.
All parameters $m_i$, $d_i$, and $k_i$ are $1$.
The position of the $i$th mass is denoted by $z_i$. We select $x_i=z_i-z_{i-1}$ and $x_{3+i}=\dot z_i - \dot z_{i-1}$, $i=1,2,3$. Assumption~\ref{asm:domi} is satisfied for $K_{\mathrm{pre}} = \0_{1 \times 6}$ because the eigenvalues of $A$ are $-1.71 \pm 0.707j$, $-1.00 \pm 1.00j$, and $-0.293\pm 0.707j$. The static nonlinearity is $\varphi (u) = \tanh(u)$, and thus $\overline{\theta}_1=1$.

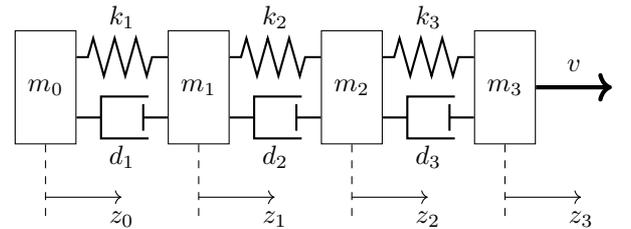
\begin{figure}[h]
\centering
\begin{tikzpicture}[scale=1,
  spring/.style={thick,decorate,decoration={zigzag,pre length=0.15cm,post length=0.15cm,segment length=8, amplitude=2.5mm}},
  mass/.style={draw, minimum width=0.8cm, minimum height=1.5cm},
  damper/.style={thick, decoration={markings, mark connection node=dmp, mark=at position 0.5 with
          {\node (dmp) [thick,inner sep=0pt,transform shape,rotate=-90,minimum width=15pt,minimum height=15pt,draw=none] {};
            \draw [thick] ($(dmp.north east)+(2pt,0)$) -- (dmp.south east) -- (dmp.south west) -- ($(dmp.north west)+(2pt,0)$);
            \draw [thick] ($(dmp.north)+(0,-5pt)$) -- ($(dmp.north)+(0,5pt)$);}}, decorate}]

  \node[mass] (m0) at (0,0) {$m_0$};
  \node[mass, right=1.2cm of m0] (m1) {$m_1$};
  \node[mass, right=1.2cm of m1] (m2) {$m_2$};
  \node[mass, right=1.2cm of m2] (m3) {$m_3$};

  \draw[spring] ([yshift=.4cm]m0.east) -- ([yshift=.4cm]m1.west) node[midway, above, yshift=7pt] {$k_1$};
  \draw[spring] ([yshift=.4cm]m1.east) -- ([yshift=.4cm]m2.west) node[midway, above, yshift=7pt] {$k_2$};
  \draw[spring] ([yshift=.4cm]m2.east) -- ([yshift=.4cm]m3.west) node[midway, above, yshift=7pt] {$k_3$};

  \draw[damper] ([yshift=-.4cm]m0.east) -- ([yshift=-.4cm]m1.west) node[midway, below, yshift=-7pt] {$d_1$};
  \draw[damper] ([yshift=-.4cm]m1.east) -- ([yshift=-.4cm]m2.west) node[midway, below, yshift=-7pt] {$d_2$};
  \draw[damper] ([yshift=-.4cm]m2.east) -- ([yshift=-.4cm]m3.west) node[midway, below, yshift=-7pt] {$d_3$};

  \draw[->, ultra thick] (m3.east) -- ++ (1,0) node[midway, above, yshift=2pt] {$v$};

  \foreach \i in {0,1,2,3} {
    \draw[dashed] (m\i.south) -- ++(0,-1) coordinate (m\i bottom);
    \draw[->] ($ (m\i.south)!0.7!(m\i bottom) $) -- ++(1,0) node[below, yshift=-2pt] {$z_\i$};
  }
\end{tikzpicture}

\caption{Mass-spring-damper system}
\label{fig:mdk_linear}
\end{figure}

For $\gamma = 0.6$, a feasible set of solutions to \eqref{eq:domi_sfb2_SI} gives 
\begin{align*}
K 
= 
\begin{bmatrix}
0.490 & -0.812 & 0.668 & 0.840 & -0.323 & 2.08
\end{bmatrix}
\end{align*}
and $\bar \varepsilon = 0.01$.
The eigenvalues of $X$ are
$-0.222$, $-0.130$, $0.0802$, $0.165$, $0.348$, and $1.73$,
and thus its inertia is $2$.
In addition, the eigenvalues of $A + B K$ are 
$-1.49+0.858j$, $-0.703 \pm 1.03j$, and $0.236 \pm 0.790j$.

Finally, it follows that $K A^{-2} B = -1.39 \neq 0$.
From Corollary~\ref{cor2:sfb}, the designed state feedback $v = \tanh (K x)$ achieves closed-loop oscillation as shown in Fig.~\ref{fig:mech_state}.
\red
\end{secex}



\section{Output Feedback and Separation Principle}
\label{sec:output_feedback}
We consider the design of a dynamic output feedback controller
for closed-loop oscillation. In~\cite{SKW:23}, an output feedback controller is derived through the solution of two coupled LMIs.
In contrast, we illustrate here a \emph{separation principle} for $p$-dominance, leading to the design of a dynamic controller for oscillations given by the combination of a state-observer and a state-feebdack controller.

The dynamic controller is given by \eqref{eq:linear}-\eqref{eq:obs_dyn}
where $K \in \bR^{m \times n}$ and $L \in \bR^{n \times q}$ are design parameters. As in classical stabilization of LTI systems, the controller~\eqref{eq:linear}-\eqref{eq:obs_dyn} for closed-loop oscillations can be constructed by separately designing $K$ and $L$.

\begin{secthm}\label{thm:ofb}
Under Assumption~\ref{asm:domi}, suppose that
\begin{enumerate}
\item there exist $\bar \varepsilon > 0$, $\bar K \in \bR^{m \times n}$, $\bar X \in \bS_n$, and $\bar \lambda:\bR^m \to [0, \gamma]$ with $\bar \lambda (\0_m) = 0$ such that~\eqref{eq:domi_sfb} and~\eqref{eq:rank_sfb} hold for all $\bar u \in \bR^m$;
\item there exist $\kappa > 0$, $\bar L \in \bR^{n \times p}$, and $Y \succ \0_{n \times n}$ such that
\begin{align}\label{eq:obs}
\He{Y (A + \gamma \I_n) + \bar L C } + \kappa \I_n \preceq \0_{n \times n}
\end{align}
holds.
\end{enumerate}
Then, $K := \bar{K}\bar{X}^{-1}$ and $L := \bar{Y}^{-1}\bar{L}$ guarantee that
almost all trajectories of the closed-loop system \eqref{eq:linear}-\eqref{eq:obs_dyn} converge to some periodic orbit.
\end{secthm}

\begin{pf}
The proof is in Appendix~\ref{app:thm:ofb}.
\qed
\end{pf}

\begin{seccor}\label{cor:ofb}
Under Assumption~\ref{asm:domi}, suppose that $\partial \varphi_i (0) = \overline{\theta}_i$, $i=1,\dots,m$,
\begin{enumerate}
\item there exist $\bar \varepsilon > 0$, $\bar K \in \bR^{m \times n}$, $\bar X \in \bS_n$, and $\hat \lambda:\{0, \overline{\theta}_1\} \times \cdots \times  \{0, \overline{\theta}_m\}\to [0, \gamma]$ with $\hat \lambda (\overline{\theta}_1,\dots,\overline{\theta}_m) = 0$ such that~\eqref{eq:domi_sfb2} holds for all $\theta \in \{0, \overline{\theta}_1\} \times \cdots \times  \{0, \overline{\theta}_m\}$ and~\eqref{eq:rank_sfb2} holds for all $\theta \in [0, \overline{\theta}_1] \times \cdots \times  [0, \overline{\theta}_m]$;
\item there exist $\kappa > 0$, $\bar L \in \bR^{n \times p}$, and $Y \succ \0_{n \times n}$ such that~\eqref{eq:obs} holds.
\end{enumerate}
Then, $K := \bar{K}\bar{X}^{-1}$ and $L := \bar{Y}^{-1}\bar{L}$ guarantee that
almost all trajectories of the closed-loop system \eqref{eq:linear}-\eqref{eq:obs_dyn} converge to some periodic orbit.
\end{seccor}

\begin{pf}
The proof follows from that of Corollary~\ref{cor:sfb} and Theorem~\ref{thm:ofb}.
\qed
\end{pf}

In the SI case, we can also extend Corollary~\ref{cor2:sfb} to the output-feedback case. In comparison to \cite{SKW:23}, 
in Theorem~\ref{thm:ofb} (resp. Corollary~\ref{cor:ofb}), LMIs~\eqref{eq:domi_sfb} (resp.~\eqref{eq:domi_sfb2}) and \eqref{eq:obs} for observer are decoupled. That is, $L$ does not depends on $K$, and vice-versa. This shows how the classical separation principle for the stabilization of LTI systems can naturally be generalized to Lur'e oscillation. 

A difference from standard observer design is that stability of $A + \gamma \I_n + L C$ is required instead of that of $A + LC$. This is related to the dominance rate $\lambda$. If $(A, C)$ is observable, this constraint is always feasible. For example, it is easy to replace the state feedback controller designed of Example~\ref{ex:mech} with an output feedback controller by additionally solving the LMI~\eqref{eq:obs}.





\section{Multi-Agent Rhythm Synchronization}
\label{sec:multiagent}
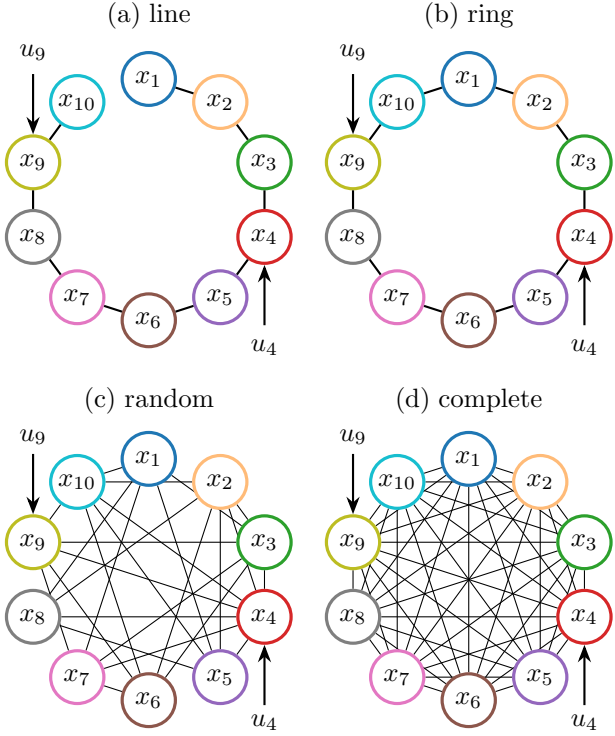
\begin{figure}[bhtp]
\begin{center}
\begin{tabular}{cc}
(a) line & (b) ring \\
\begin{tikzpicture}[line width=4pt]
  \foreach \i in {1,...,10} {
  \pgfmathtruncatemacro{\colorindex}{\i}
    \node (P\i) at ({90 - 36*(\i-1)}:1.6) 
      [circle, draw=c\colorindex, very thick, inner sep=.4mm, minimum size=7mm] 
      {$x_{\i}$};
  }
  
  \foreach \i in {1,...,9} {
    \pgfmathtruncatemacro{\j}{\i + 1}
    \draw[thick] (P\i) -- (P\j);
  }

  \node (u4) [below=8mm of P4] {$u_4$};
  \draw[thick, ->, >={Stealth}] (u4) -- (P4);
  
  \node (u9) [above=8mm of P9] {$u_9$};
  \draw[thick, ->, >={Stealth}] (u9) -- (P9);
  \end{tikzpicture}
&
\begin{tikzpicture}[line width=4pt]
  \foreach \i in {1,...,10} {
  \pgfmathtruncatemacro{\colorindex}{\i}
    \node (P\i) at ({90 - 36*(\i-1)}:1.6) 
      [circle, draw=c\colorindex, very thick, inner sep=.4mm, minimum size=7mm] 
      {$x_{\i}$};
  }

  \foreach \i in {1,...,9} {
    \pgfmathtruncatemacro{\j}{\i + 1}
    \draw[thick] (P\i) -- (P\j);
  }
  \draw[thick] (P10) -- (P1);
  
  \node (u4) [below=8mm of P4] {$u_4$};
  \draw[thick, ->, >={Stealth}] (u4) -- (P4);
  
  \node (u9) [above=8mm of P9] {$u_9$};
  \draw[thick, ->, >={Stealth}] (u9) -- (P9);
\end{tikzpicture}\\[0mm]
(c) random & (d) complete \\
\begin{tikzpicture}[on background layer/.style={preaction={draw=white, line width=4pt}}]
   \foreach \i in {1,...,10} {
  \pgfmathtruncatemacro{\colorindex}{\i}
    \node (P\i) at ({90 - 36*(\i-1)}:1.6) 
      [circle, draw=c\colorindex, fill=white, very thick, inner sep=.4mm, minimum size=7mm] 
      {$x_{\i}$};
  }

  \begin{pgfonlayer}{background}
    \foreach \i/\j in {1/3,1/5,1/7,1/8,1/10,
                       2/3,2/4,2/5,2/6,2/8,2/10,
                       3/4,3/6,3/7,3/9,
                       4/5,4/7,4/8,4/9,4/10,
                       5/8,5/10,
                       6/7,6/9,6/10,
                       7/9,
                       9/10} {
      \draw (P\i) -- (P\j);
    }
  \end{pgfonlayer}

  \node (u4) [below=8mm of P4] {$u_4$};
  \draw[thick, ->, >={Stealth}] (u4) -- (P4);
  
  \node (u9) [above=8mm of P9] {$u_9$};
  \draw[thick, ->, >={Stealth}] (u9) -- (P9);
\end{tikzpicture}
&
\begin{tikzpicture}[on background layer/.style={preaction={draw=white, line width=4pt}}]
  \foreach \i in {1,...,10} {
    \pgfmathtruncatemacro{\colorindex}{\i}
    \node (P\i) at ({90 - 36*(\i-1)}:1.6) 
      [circle, draw=c\colorindex, fill=white, very thick, inner sep=.4mm, minimum size=7mm] 
      {$x_{\i}$};
  }
  \begin{pgfonlayer}{background}
  \foreach \i in {1,...,10} {
    \foreach \j in {1,...,10} {
      \ifnum\i<\j
        \draw (P\i) -- (P\j);
      \fi
    }
  }
  \end{pgfonlayer}

  \node (u4) [below=8mm of P4] {$u_4$};
  \draw[thick, ->, >={Stealth}] (u4) -- (P4);
  
  \node (u9) [above=8mm of P9] {$u_9$};
  \draw[thick, ->, >={Stealth}] (u9) -- (P9);
  \end{tikzpicture}
\end{tabular}
\end{center}

\caption{Network topologies}
\label{fig:toplogy}
\end{figure}

We demonstrate the potential of our proposed theoretical conditions with a new control design for synchronization in multi-agent systems. Rather than the classical consensus approach, our method guarantees that agents phase-lock to a common rhythm. We designed a controller for a finite number of agents with heterogeneous dynamics. At a steady state, the relative phase between any two agents is locked, even when they do not converge to a uniform behavior (as they would in consensus). We consider two scenarios based the four network topologies of Fig. \ref{fig:toplogy}. Scenario A models agents as first-order lags, while Scenario B considers simple electro-mechanical systems.

\begin{figure*}[htbp]
\begin{center}
\begin{subfigure}{0.47\textwidth}
\includegraphics[width=\columnwidth]{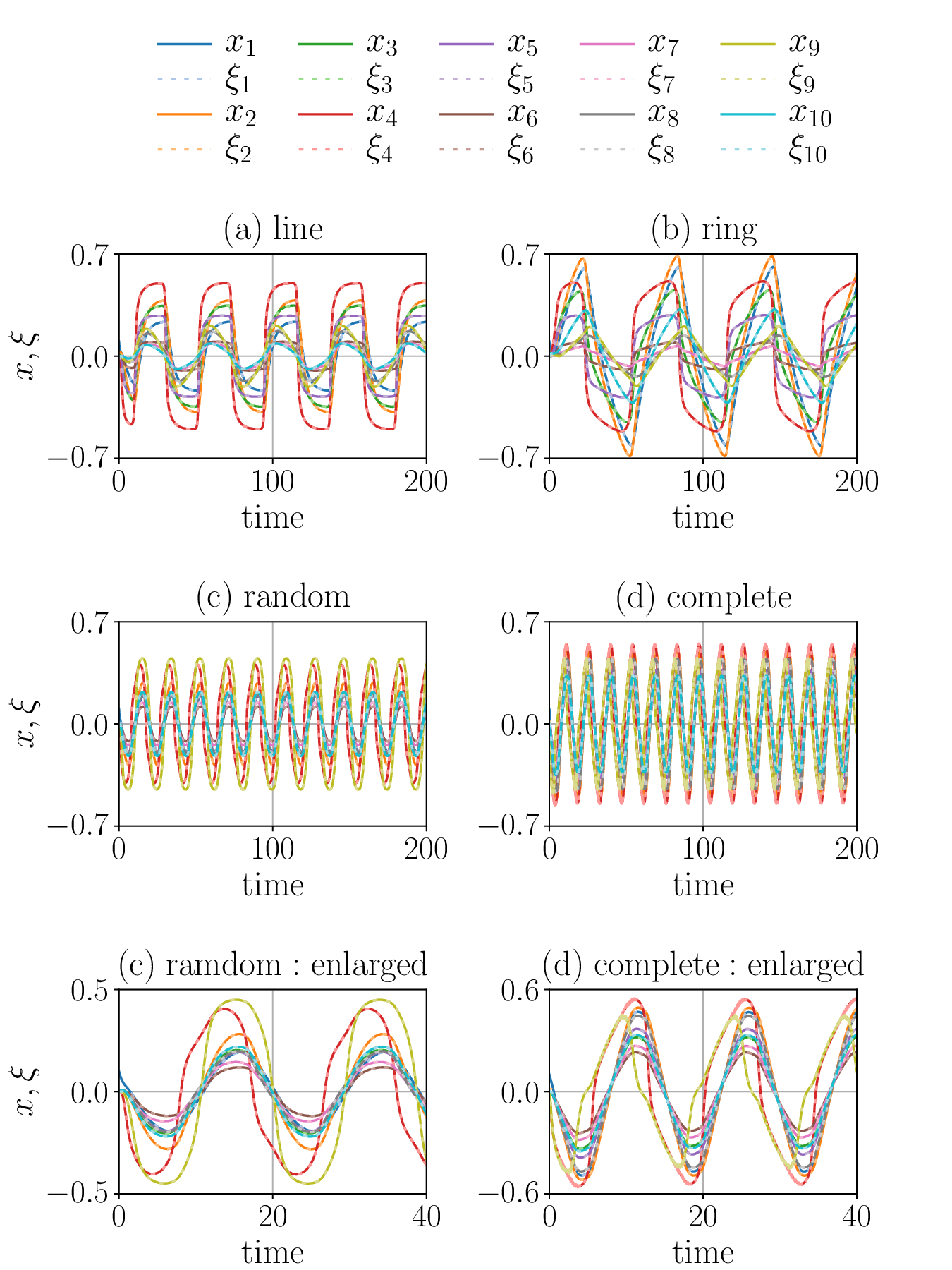}
\caption{First order lags}
\label{fig:agent}
\end{subfigure}
\begin{subfigure}{0.47\textwidth}
\includegraphics[width=\columnwidth]{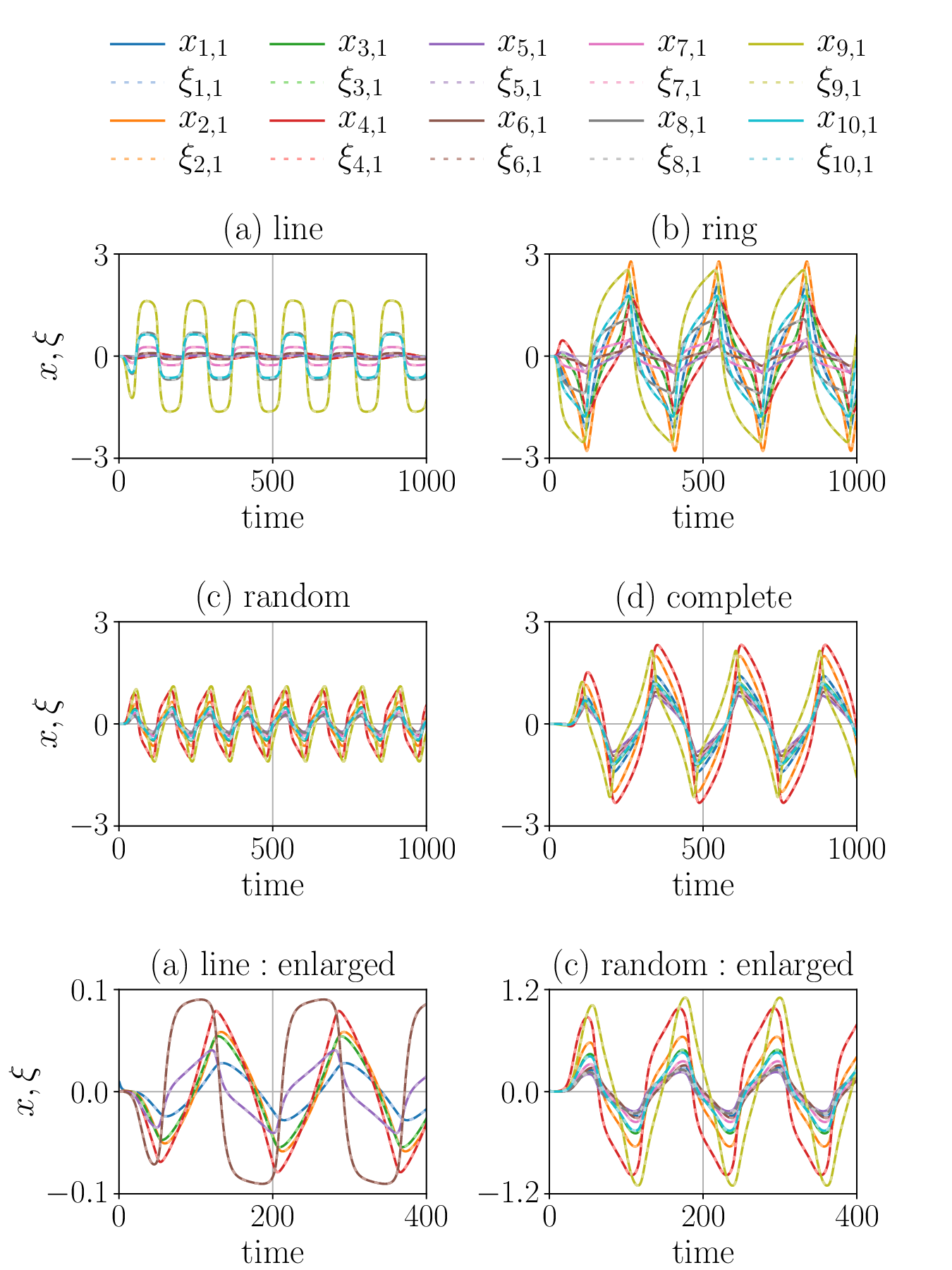}
\caption{Electro-mechanical systems}
\label{fig:agent2}
\end{subfigure}
\caption{Multi-agent rhythm synchronization}
\label{fig:agent0}
\end{center}
\end{figure*}



\subsection{First order lags}
Consider a multi-agent system with $10$ agents 
\begin{align*}
\dot x_i &= - a_i x_i + \sum_{k \in \mathcal{N}_i} W_{i,k} x_k  + \varphi(u_i), & i &\in \{4, 9\}\\
\dot x_i &= - a_i x_i + \sum_{k \in \mathcal{N}_i} W_{i,k} x_k, & i &\not\in \{4, 9\}
\end{align*}

where $x_i \in \bR$, $i \in \{1,\dots,10\}$, $u_4, u_9 \in \bR$, and $\varphi(u) = \tanh(u)$. 
Each $a_i$ is randomly generated between $1$ and $5$, and the resulting values are
\begin{align*}
a = \begin{bmatrix} 1.62 & 1.52 & 2.54 & 3.24 & 2.15 & 3.62 & 3.07 & 1.72 & 2.95 & 2.42 \end{bmatrix}.
\end{align*}
Each set $\mathcal{N}_i$ contains the indices of the nodes connected to the node $i$, as defined by the topologies in Figure \ref{fig:toplogy}. 
We consider networks characterized by undirected graphs with the following topologies: (a) line with wights $W_{i,k} = 1$, (b) ring with $W_{i,k} = 1$, (c) random with $W_{i,k} = 1/3$, and (d) complete with $W_{i,k} = 1/4$. The weights for (c) random and (d) complete are selected such that the network system is Hurwitz. 

For each topology, we consider the same output
\begin{align}
y_1 = x_4 \qquad y_2  = x_9. \label{eq:multiagent_output}
\end{align}
Synchronization to a common rhythm is achieved by designing an output feedback controller~\eqref{eq:linear}-\eqref{eq:obs_dyn} for closed-loop oscillations.  Assumption~\ref{asm:domi} holds for $K_{\mathrm{pre}}=\0_{2 \time 10}$ for each topology. We design $K$ and $L$ by separately solving~\eqref{eq:domi_sfb2} and~\eqref{eq:obs}, where $\overline{\theta}_1=\overline{\theta}_2 = 1$. Each parameter is selected as in Table~\ref{tab:para}. 
For each topology, the phase-locked solutions of the controlled 
closed-loop multi-agent system are illustrated in Figure~\ref{fig:agent}.

\begin{table}[htbp]
    \centering
    \scalebox{0.8}{
\begin{tabular}{|l||c|c|c|c|c|}\hline
Topology &$\!\!\gamma\!\!$ & $\!\!\hat \lambda (1,0)\!\!$ & $\!\!\hat \lambda (0,1)\!\!$ & $\!\!\overline{\varepsilon}\!\!$ & $\!\!\kappa \times 10^{-4}\!\!$ \\ \hline \hline
(a) line & $\!\!0.874\!\!$ & $\!\!0.05\gamma\!\!$ & $\!\!0.75\gamma\!\!$ & $\!\!0.10\!\!$ & $\!\!0.975\!\!$\\ \hline
(b) ring & $\!\!0.825\!\!$ & $\!\!0.05\gamma\!\!$ & $0.75\gamma$ & $\!\!0.10\!\!$ & $\!\!1\!\!$ \\ \hline
(c) random & $\!\!1.61\!\!$ & $\!\!0.9\gamma\!\!$ & $0.9\gamma$ & $\!\!0.10\!\!$ & $\!\!1\!\!$ \\ \hline
(d) complete & $\!\!1.87\!\!$ & $\!\!0.4\gamma\!\!$ & $0.8\gamma$ & $\!\!0.08\!\!$ & $\!\!1\!\!$ \\ \hline
\end{tabular}
}
\caption{}
\label{tab:para}
\end{table}
\vspace{-4mm}





\subsection{Linear electro-mechanical systems}
\vspace{-2mm}
We replace each single-integrator agent by a electro-mechanical system
\begin{align*}
\dot x_i &= - A_i x_i + B \left(\sum_{k \in \mathcal{N}_i} W_{i,k} C x_k  + \varphi(u_i)\right), \quad i \in \{4, 9\} \\
\dot x_i &= - A_i x_i + B \left(\sum_{k \in \mathcal{N}_i} W_{i,k} C x_k\right), \hspace{16mm}  i \not\in \{4, 9\}\\
& A_i = -\begin{bmatrix}
0 & 1 & 0\\
-1 & -b_i & 1\\
0 & -1 & -c_i
\end{bmatrix},
\quad 
B = \begin{bmatrix}
0 \\ 0\\ 1
\end{bmatrix},
\quad
C = \begin{bmatrix}
1 & 0 & 0
\end{bmatrix},
\end{align*}
where $x_i = [\begin{matrix}x_{i,1} & x_{i,2} & x_{i,3} \end{matrix}]^\top \in \bR^3$ with position $x_{i,1}$ and velocity $x_{i,2}$ of mass $i$ and current $x_{i,3}$ of motor $i$ and $u_4, u_9 \in \bR$. Each $b_i$ and $c_i$ are randomly generated between $1$ and $3$ and between $1$ and $4$, respectively, and the resulting values are
\begin{align*} 
b &= \begin{bmatrix} 2.59 & 1.75 & 1.61 & 2.36 & 1.16 & 2.95 & 2.13 & 2.49 & 2.86 & 1.22 \end{bmatrix}\\
c &= \begin{bmatrix} 2.07 & 1.39 & 2.48 & 1.90 & 3.79 & 3.25 & 2.92 & 2.74 & 1.42 & 2.56 \end{bmatrix}.
\end{align*}
The function $\varphi(\cdot)$ and graph topologies including the weights $W_{i,k}$ are the same as the single-integrator case. 

For each topology, we consider the same output corresponding to \eqref{eq:multiagent_output}
\begin{align}
y_1 = C x_4 \qquad y_2  = C x_9,
\end{align}
and Assumption~\ref{asm:domi} holds for $K_{\mathrm{pre}}=\0_{2 \times 30}$. As in the previous case, we derive an output feedback controller~\eqref{eq:linear}-\eqref{eq:obs_dyn} for closed-loop oscillation. Each parameter is selected as in Table~\ref{tab:para2}. 
Controlled synchronization to a common rhythm is illustrated in Figure~\ref{fig:agent2}.

\begin{table}[htbp]
    \centering
    \scalebox{0.8}{
\begin{tabular}{|l||c|c|c|c|c|}\hline
Topology &$\!\!\gamma\!\!$ & $\!\!\hat \lambda (1,0)\!\!$ & $\!\!\hat \lambda (0,1)\!\!$ & $\!\!\overline{\varepsilon}\!\!$ & $\!\!\kappa\!\!$ \\ \hline \hline
(a) line & $\!\!0.228\!\!$ & $\!\!\gamma\!\!$ & $\!\!0.25\gamma\!\!$ & $\!\!0.1\!\!$ & $\!\!10^{-4}\!\!$\\ \hline
(b) ring & $\!\!0.144\!\!$ & $\!\!0.7\gamma\!\!$ & $\!\!0.2\gamma\!\!$ & $\!\!0.1\!\!$ & $\!\!10^{-4}\!\!$ \\ \hline
(c) random & $\!\!0.321\!\!$ & $\!\!0.65\gamma\!\!$ & $\!\!0.9\gamma\!\!$ & $\!\!0.1\!\!$ & $\!\!10^{-4}\!\!$ \\ \hline
(d) complete & $\!\!0.214\!\!$ & $\!\!0.8\gamma\!\!$ & $\!\!0.2\gamma\!\!$ & $\!\!0.1\!\!$ & $\!\!10^{-4}\!\!$ \\ \hline
\end{tabular}
}
\caption{}
\label{tab:para2}
\end{table}





\section{Open Challenges in Nonlinear Systems}\label{sec:nonlinear}

In this section, we discuss the extension to 
nonlinear systems of the form $\dot x = f(x) + B \varphi(u)$,
where $f: \bR^n \to \bR^n$ is of class $C^1$. For the sake of simplicity, we assume that the origin is an unforced equilibrium point, i.e., $f(\0_n) = \0_n$. The closed-loop system with the state feedback~\eqref{eq:static_u_state_feedback} is
\begin{align}\label{eq:ncls}
\dot x = f(x) + B \varphi (Kx).
\end{align}

\begin{secasm}\label{asm:ndomi}
For the nonlinear system \eqref{eq:ncls}, $\dot x = f(x)$ is strictly $0$-dominant with respect to $\gamma_0 \ge 0$ and strictly $2$-dominant with respect to some rate $\gamma_2 > 0$.
\red
\end{secasm}

\begin{secthm}\label{thm:nsfb}
Under Assumption~\ref{asm:ndomi}, let $\bar \varepsilon > 0$, $\bar K \in \bR^{m \times n}$, $\bar X \in \bS_n$, and $\bar \lambda:\bR^m \to [0, \gamma]$ with $\bar \lambda (\0_m) = 0$ 
be such that
\begin{subequations}\label{eq:domi_sfb_non}
\begin{align}
\He{(\partial f(x) + \gamma_2 \I_n) \bar X} + \bar \varepsilon  \I_n &\preceq \0_{n\times n} \label{eq1:domi_sfb_non}\\
{\rm He} \big((\partial f(x) + \bar \lambda (\bar u) \I_n) \bar X \hspace{10mm} & \nonumber\\
 + B \partial \varphi (\bar u) \bar K \big) + \bar \varepsilon  \I_n &\preceq \0_{n\times n} \label{eq2:domi_sfb_non}\\
{\rm rank}\left( \int_0^1 \partial f(s x) ds + B \partial \varphi(\bar u) K \right) &= n
\label{eq:rank_sfb_non}
\end{align}
\end{subequations}
hold for all $x \in \bR^n$ and $\bar u \in \bR^m$.
Then, $K := \bar{K}\bar{X}^{-1}$ guarantees that almost all trajectories of the closed-loop system~\eqref{eq:ncls} converge to some periodic orbit.
\end{secthm}
\begin{pf}
The proof is in Appendix~\ref{app:thm:nsfb}.
\qed
\end{pf}

While Assumption~\ref{asm:ndomi} and Theorem~\ref{thm:nsfb} are straightforward extensions
of Assumption \ref{asm:domi} and Theorem \ref{thm:sfb}, respectively, 
their practical verification \emph{for all} $x$ is a much harder problem.
In special cases, this can be reduced to a finite set of LMIs, via convex relaxation.
An example is shown below.

We make no use of the pre-compensator $K_{\mathrm{pre}}$ in the nonlinear case. 
The main role of pre-compensator is to guarantee Assumption
\ref{asm:domi} via pole placement. However, an equivalent approach that holds for the Jacobian $\partial f(x)$ uniformly in $x$ is currently not available. 
Furthermore, even in the SI case, verifying the rank condition~\eqref{eq:rank_sfb_non} is extremely challenging, even numerically. This requires the numerical integration of $\partial f(s x)$ with respect to $s$. 

Finally, in the nonlinear case, we cannot take advantage of the separation principle for the output feedback controller. This case is discussed in \cite{SKW:23}, which
proposes \emph{coupled} conditions for state-feedback and observer.

\begin{figure}[htbp]
\centering
\includegraphics[width=0.9\columnwidth]{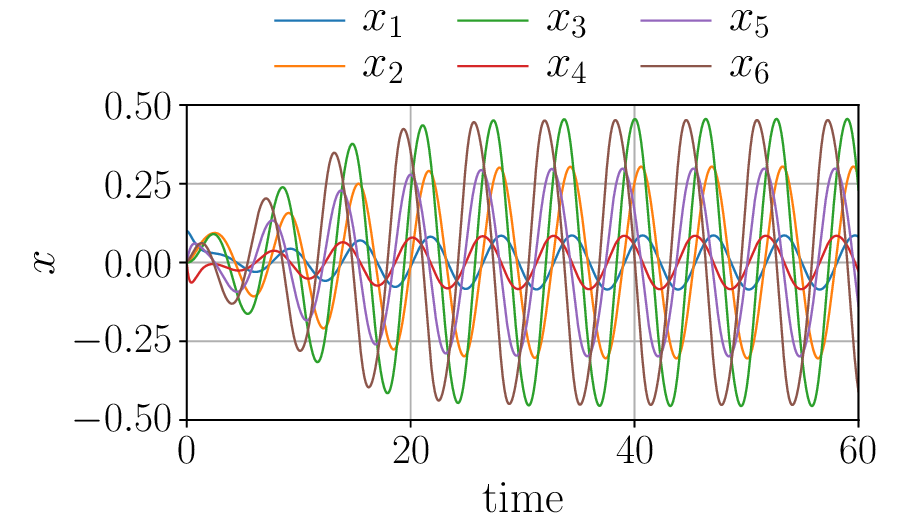}
\caption{Solution of simplified nonlinear robotics model}
\label{fig:sim_snake2}
\end{figure}

\begin{secex}
We consider again the mass-spring-damper system 
of Fig.~\ref{fig:mdk_linear}. We use the same notations and parameters as Example~\ref{ex:mech} except for the tail; $k_1 (z_1-z_0)$ is replaced with nonlinear spring $k_1 (z_1-z_0) + \tanh (z_1 - z_0)$ representing a position-dependent force.

For each $x \in \bR^6$, there exists $\eta (x) \in [0,1]$ such that $\partial f(x) = \eta(x) A_1 + (1-\eta (x)) A_2$, where
\begin{align*}
&A_i
=
\begin{bmatrix}
\0_{3 \times 3} &\I_3\\
\bar A_i & \bar A_1
\end{bmatrix}, \quad 
\bar A_i =
\begin{bmatrix}
a_i & 1 & 0 \\
b_i & -2 & 1 \\
0 & 1 & -2
\end{bmatrix},
\quad i = 1,2\\
&\quad a_1 = -2, \; b_1 = 1, \; a_2 = b_2 = 0.
\end{align*}
Assumption~\ref{asm:ndomi} and the conditions \eqref{eq1:domi_sfb_non},\eqref{eq2:domi_sfb_non} can be verified by using $A_1$ and $A_2$. For instance, for  Assumption~\ref{asm:ndomi}, given $\gamma_0, \gamma_2 \ge 0$, there exist $\varepsilon_0, \varepsilon_2 > 0$, $X_0 \succ \0_{6 \times 6}$, and $X_2 \in \bS_6$ of inertia $2$ such that
\begin{align*}
\He{X_j (A_i + \gamma_j \I_n)} + \varepsilon_j  \I_n \preceq \0_{6 \times 6}, \quad i=1,2
\end{align*}
hold for each $j=0,2$. For $\gamma_0 = 0$, $\gamma_2 = 0.7$, and $\varepsilon_0 = \varepsilon_2 = 0.1$, the sets of eigenvalues of $X_0$ and $X_2$ satisfying these are $0.0361$, $0.144$, $0.395$, $0.660$, $3.20$, $4.91$ and $-6.93$, $-2.64$, $0.590$, $3.00$, $7.82$, $15.9$, respectively. Thus, this system satisfies Assumption~\ref{asm:ndomi}.


Next, we design Lur'e state feedback $u = \tanh (Kx)$ for closed-loop oscillation. Adapting Corollary~\ref{cor:sfb}, it is possible to show that if there exist $\bar \varepsilon > 0$, $\bar K \in \bR^{1 \times 6}$, and $\bar X \in \bS_6$ such that
\begin{align*}
\He{(A_i + \gamma_2 \I_n) \bar X} + \bar \varepsilon  \I_n &\preceq \0_{6 \times 6}, \quad i= 1,2 \\
\He{A_i \bar X + B \bar K} + \bar \varepsilon \I_n & \preceq \0_{6 \times 6}, \quad i= 1,2.
\end{align*}
hold, then they satisfy~\eqref{eq1:domi_sfb_non},\eqref{eq2:domi_sfb_non} for $\bar \lambda (\bar u) \equiv 0$. Solving this for $\bar \varepsilon = 0.1$ gives 
\begin{align*}
K = 
\begin{bmatrix}
0.435 & -0.547 & 0.198 & 0.491 & 0.235 & 1.88
\end{bmatrix}.
\end{align*}
The eigenvalues of $X$ are $-0.259$, $-0.186$, $0.0762$, $0.130$, $0.408$, $0.819$, and thus its inertia is $2$.
This leads to closed-loop oscillation, as illustrated
in Fig.~\ref{fig:sim_snake2}.
\red
\end{secex}


\section{Conclusion}



We have developed an LMI framework for designing dynamic output-feedback controllers for Lur’e oscillations in the MIMO setting by extending dominant system theory to state-dependent rates. We have first derived conditions for state-feedback design, and then proposed a separation principle. As a result, a dynamic output-feedback Lur’e oscillator can be implemented by additionally designing a standard observer for LTI systems shifted by the dominance rate.

This paper is a first step toward developing a numerically tractable approach for designing Lur’e oscillators in the MIMO setting. However, several open problems remain. For instance, establishing conditions that guarantee the uniqueness of the closed-loop equilibrium point is still unresolved. In the SISO case, this issue can be addressed by applying a small perturbation to the state-feedback gain. Extension to nonlinear plants also remains open questions, as mentioned in Section~\ref{sec:nonlinear}. 

The proposed Lur’e oscillation synthesis has been applied
to a new control design for synchronization in multi-agent systems. new control design for synchronization in multi-agent systems. For various graph topologies with heterogeneous agents, rhythmic synchronization have been achieved. However, the mechanisms determining the resulting oscillation frequency and the shape of the limit cycle are not yet understood. Other interesting open problems include robustness analysis of oscillations and the development of an interconnection theory. 



\appendix
\section{Proof of Theorem~\ref{thm:sfb}}\label{app:thm:sfb}

We apply Proposition~\ref{prop:domi}.
By Assumption~\ref{asm:domi}, the LTI system \eqref{eq:linear} is bounded-input and bounded-state (BIBS) stable.
Thus, all closed-loop trajectories are bounded by the boundedness of $\varphi$.

By Assumption~\ref{asm:domi}, $\bar X$ satisfying~\eqref{eq1:domi_sfb} has inertia $2$. Noting that inversion preserves the inertia, define $X := \bar X^{-1}$ and $\lambda (x) := \bar \lambda (Kx)$, $x \in \bR^n$. Also, let $\varepsilon > 0$ be such that $\varepsilon \I_n \preceq \bar \varepsilon X^2$. Then, multiplying \eqref{eq2:domi_sfb} by $X$ from both sides and substituting $\bar u = K x$ yield
\begin{align}\label{pf1:domi_sfb}
&\He{X (A_{\mathrm{pre}} + \lambda (x) \I_n + B \partial \varphi (K x) K)} + \varepsilon  \I_n \preceq \0_{n \times n} 
\end{align}
for all $x \in \bR^n$.

From~\eqref{pf1:domi_sfb}, the closed-loop system is strictly $2$-dominant with rate $\lambda (x)$. Thus, $\lambda (\0_n) = \bar \lambda (\0_m) = 0$ implies that $A+B \partial \varphi (\0_m) K$ has $2$ positive real eigenvalues. Since this is the system matrix of the linearized closed-loop system at the origin, the closed-loop system is unstable at the origin. 

Finally, we show that the origin is the unique equilibrium.
Because of $\varphi (\0_m) = \0_m$, $x^* \in \bR^n$ is an equilibrium of the closed-loop system if and only if
\begin{align}\label{pf2:domi_sfb}
\0_n &=  A_{\mathrm{pre}} x^* + B \varphi (K x^*) \nonumber\\
&= \left( A_{\mathrm{pre}} + B \int_0^1 \partial \varphi (s K x^*) ds K \right) x^*.
\end{align}
From the diagonal structure~\eqref{eq:diagonal} of $\varphi$, we can apply the mean value theorem component-wise. Namely, for each $x^* \in \bR^n$, there exists $\bar u \in \bR^m$ such that
\begin{align*}
\int_0^1 \partial \varphi (s K x^*) ds = \partial \varphi (\bar u).
\end{align*}
From \eqref{eq:rank_sfb} and non-singularity of $\bar X$, \eqref{pf2:domi_sfb} has a unique solution that is $x^* = \0_n$.
\qed



\section{Proof of Corollary~\ref{cor:sfb}}\label{app:cor:sfb}
First, \eqref{eq1:domi_sfb2} is nothing but \eqref{eq1:domi_sfb}. We show that \eqref{eq2:domi_sfb2} implies \eqref{eq2:domi_sfb}.
Let $\theta^j$, $j=1,\dots,2^m$ denote each distinct element of $\{0, \overline{\theta}_1\} \times \cdots \times  \{0, \overline{\theta}_m\}$. From~\eqref{eq:bound}, $\partial \varphi (\bar u)$ takes a value in $[0, \overline{\theta}_1] \times \cdots \times [0, \overline{\theta}_m]$. Thus, for each $\bar u \in \bR^m$, there exist $\alpha^j(\bar u) \ge 0$, $j=1,\dots,2^m$ such that
\begin{align*}
\sum_{j=1}^{2m} \alpha^j(\bar u) {\rm diag}(\theta^j) = \partial \varphi (\bar u), \quad
\sum_{j=1}^{2m} \alpha^j(\bar u) = 1.
\end{align*}
Also, define
\begin{align*}
\bar \lambda (\bar u) := \sum_{j=1}^{2m} \alpha^j(\bar u) \hat \lambda (\theta^j) \ge 0,
\quad
\forall \bar u \in \bR^m.
\end{align*}
Then, it follows from~\eqref{eq2:domi_sfb2} that
\begin{align*}
&\He{(A_{\mathrm{pre}} + \bar \lambda (\bar u) \I_n) \bar X + B \partial \varphi (\bar u) \bar K} + \bar \varepsilon  \I_n \\
&= \sum_{j=1}^{2m} \alpha^j(\bar u) \He{ (A_{\mathrm{pre}} + \hat \lambda (\theta^j)) \bar X + B {\rm diag}(\theta^j) \bar K} \\
&\qquad + \bar \varepsilon  \I_n \preceq \0_{n \times n}.
\end{align*}
Therefore, \eqref{eq2:domi_sfb} holds. Especially when $\bar u = \0_m$, we have 
\begin{align*}
&\He{(A_{\mathrm{pre}} \bar X + B \partial \varphi(\0_m) \bar K} + \bar \varepsilon  \I_n\\
&=\He{(A_{\mathrm{pre}} + \hat \lambda  (\bar \theta_1,\dots,\bar \theta_m)) \bar X + B {\rm diag} (\bar \theta_1,\dots,\bar \theta_m) \bar K}\nonumber\\
&\qquad + \bar \varepsilon  \I_n \preceq \0_{n \times n}.
\end{align*}
This implies that the closed-loop system is unstable at the origin.
Finally, since $\partial \varphi (\bar u) \in [0, \overline{\theta}_1] \times \cdots \times [0, \overline{\theta}_m]$ for all $\bar u \in \bR^m$, \eqref{eq:rank_sfb2} implies \eqref{eq:rank_sfb}.
\qed



\section{Proof of Corollary~\ref{cor2:sfb}}\label{app:cor2:sfb}
In the SI case ($m=1$), \eqref{eq:domi_sfb2} becomes \eqref{eq1:domi_sfb2_SI},\eqref{eq2:domi_sfb2_SI}. We show that $KA_{\mathrm{pre}}^{-2}B \neq 0$ implies~\eqref{eq:rank_sfb2} by contraposition. The rank condition~\eqref{eq:rank_sfb2} does not hold if and only if $A_{\mathrm{pre}} + \theta_1 B K$ has a zero eigenvalue for some $\theta_1 \in [0, \overline{\theta}_1]$. This is also equivalent to that the root locus of \eqref{eq:rlocus} passes the origin for some $\theta_1 \in (0, \overline{\theta}_1)$. From $2$-dominance and symmetry of the root locus with respect to the real axis, if the root locus passes the origin, then its multiplicity is exactly $2$. Namely, it follows that 
\begin{align*}
\left.\frac{d ( 1- \theta_1 K(sI-A_{\mathrm{pre}})^{-1}B)}{ds}\right|_{s=0}
= \theta_1 KA_{\mathrm{pre}}^{-2}B = 0,
\end{align*}
i.e., $KA_{\mathrm{pre}}^{-2}B = 0$.
\qed



\section{Proof of Corollary~\ref{thm:ofb}}\label{app:thm:ofb}
Introducing error $e := \xi - x$, the closed-loop system can be described as
\begin{align}\label{pf1:obs}
\left\{\begin{alignedat}{2}
\dot x &{} = A_{\mathrm{pre}} x  + B (\varphi (K (x + e)) + K_{\mathrm{pre}} e)\\
\dot e &{} = (A + L C) e
\end{alignedat}\right.
\end{align}
for $L:= Y^{-1} \bar L$.
The corresponding Jacobian matrix is
\begin{align*}
&J(x,e)
:= \\
&\begin{bmatrix}
A_{\mathrm{pre}} + B \partial \varphi (K (x + e)) K & B (\partial \varphi (K (x + e)) K + K_{\mathrm{pre}})\\
\0_{n \times n} & A + L C
\end{bmatrix}.
\end{align*}

To show strict $2$-dominance of the closed-loop system~\eqref{pf1:obs}, define the block diagonal matrix with a scaling parameter $s > 0$:
\begin{align*}
Z_s 
:= 
\begin{bmatrix}
X & \0_{n \times n} \\
\0_{n \times n} & s Y
\end{bmatrix}.
\end{align*}
Its inertia is $2$ because the inertia of $X$ and $Y$ are $2$ and $0$, respectively.

From the proof of Theorem~\ref{thm:sfb}, there exist $\varepsilon  > 0$, $K \in \bR^{m \times n}$, $X \in \bS_n$ with inertia $2$, and $\lambda: \bR^n \to [0, \gamma]$ with $\lambda (\0_n) = 0$ such that~\eqref{pf1:domi_sfb}, i.e.,
\begin{align}\label{pf3:obs}
&\He{X (A_{\mathrm{pre}} + \lambda (x+e) \I_n + B \partial \varphi (K(x+e)) K )}\nonumber\\
&\quad + \varepsilon  \I_n \preceq \0_{n \times n}
\end{align}
holds for all $x, e \in \bR^n$. Using $L = Y^{-1} \bar L$, \eqref{eq:obs} can be rewritten as
\begin{align}\label{pf4:obs}
\He{Y (A + \gamma \I_n + L C) } + \kappa \I_n \preceq \0_{n \times n}.
\end{align}

It follows from~\eqref{pf3:obs} and \eqref{pf4:obs} that
\begin{align*}
&\He{Z_s (J(x,e) + \lambda (x+e) \I_{2n})} \\
&\preceq 
\begin{bmatrix}
- \varepsilon \I_n & X B (\partial \varphi (K(x+e)) K + K_{\mathrm{pre}}) \\
* & s  (\lambda (x+e) - \gamma) Y - s \kappa \I_n
\end{bmatrix} \\
&\preceq 
\begin{bmatrix}
- \varepsilon \I_n & X B (\partial \varphi (K(x+e)) K + K_{\mathrm{pre}}) \\
* & - s\kappa \I_n
\end{bmatrix},
\end{align*}
where $\lambda (\cdot ) \le \gamma$ and $Y \succ \0_{n \times n}$ are used in the last inequality.
Since $\partial \varphi (\cdot)$ is bounded, the inequality becomes negative definite for sufficiently large $s > 0$.
Thus, the closed-loop system~\eqref{pf1:obs} is strictly $2$-dominant with rate $\lambda (x+e)$.
Also, from $\lambda (\0_n) = 0$, its linearization at the origin is unstable. 

Each trajectory of the closed-loop system~\eqref{pf1:obs} is bounded because $A_{\mathrm{pre}}$ and $A+LC$ are Hurwitz, and $\varphi$ is bounded.
It remains to show that the origin is the unique equilibrium.
Since $A + L C$ is Hurwitz, all equilibrium points of the closed-loop system~\eqref{pf1:obs} satisfy $e = \0_n$, and thus, $A_{\mathrm{pre}} x + B \varphi (K x) = \0_n$. This has a unique solution $x = \0_n$ if the rank condition~\eqref{eq:rank_sfb} holds.
\qed



\section{Proof of Corollary~\ref{thm:nsfb}}\label{app:thm:nsfb}
We apply Proposition~\ref{prop:domi}.
Strict $0$-dominance with $\gamma_0 = 0$ (i.e., contraction) and $f(\0_n)=\0_n$ imply (incremental) input-to-state stability~\cite[Corollary 3.16]{Bullo:22}. By the boundedness of $\varphi$, all closed-loop trajectories are bounded.

Define $X := \bar X^{-1}$ and $\lambda (x) := \bar \lambda (Kx)$, $x \in \bR^n$. Also, let $\varepsilon > 0$ be such that $\varepsilon \I_n \preceq \bar \varepsilon X^2$. Then, multiplying \eqref{eq1:domi_sfb_non},\eqref{eq2:domi_sfb_non} by $X$ from both sides and substituting $\bar u = K x$ yield
\begin{align*}
\He{X(\partial f(x) + \gamma_2 \I_n)} + \varepsilon \I_n &\preceq \0_{n \times n}\\
\He{X(\partial f(x) + \lambda (K x) \I_n) + B \partial \varphi (K x) K)} + \bar \varepsilon  \I_n &\preceq \0_{n\times n}
\end{align*}
for all $x \in \bR^n$. By Assumption~\ref{asm:domi} and the first inequality, $X$ has inertia $2$. From the second inequality, the closed-loop system is strictly $2$-dominant. Also, $\lambda (\0_n) = \bar \lambda (\0_m) = 0$ implies that $\partial f(\0_n) + B \partial \varphi (\0_m) K$ has $2$ positive real eigenvalues. Therefore, the linearization at the origin (an equilibrium) is unstable. 

Similarity to the proof of Theorem~\ref{thm:sfb}, one can show that the rank condition~\eqref{eq:rank_sfb_non} implies the uniqueness of an equilibrium.
\qed



\bibliographystyle{plain}
\bibliography{ref}

\begin{thebibliography}{10}

\bibitem{AVK:66}
A.~A. Andronov, A.~A. Vitt, and S.~E. Khaikin.
\newblock {\em Theory of Oscillators}.
\newblock Elsevier, 1966.

\bibitem{Astrom:95}
Karl~J Astr{\"o}m.
\newblock Oscillations in systems with relay feedback.
\newblock In {\em Adaptive Control, Filtering, and Signal Processing}, pages 1--25. Springer, 1995.

\bibitem{Bullo:22}
F.~Bullo.
\newblock {\em Contraction Theory for Dynamical Systems}.
\newblock Kindle Direct Publishing, 2022.

\bibitem{CF:24}
W.~Che and F.~Forni.
\newblock Dominant mixed feedback design for stable oscillations.
\newblock {\em IEEE Transactions on Automatic Control}, 69(2):1133--1140, 2024.

\bibitem{FS:19}
F.~Forni and R.~Sepulchre.
\newblock Differential dissipativity theory for dominance analysis.
\newblock {\em IEEE Transactions on Automatic Control}, 64(6):2340--2351, 2019.

\bibitem{GV:68}
A.~Gelb and W.~E. {Vander Velde}.
\newblock {\em Multiple-Input Describing Functions and Nonlinear System Design}.
\newblock McGraw-Hill, 1968.

\bibitem{GMD:01}
J~.~M. Goncalves, A.~Megretski, and M.~A. Dahleh.
\newblock Global stability of relay feedback systems.
\newblock {\em IEEE Transactions on Automatic Control}, 46(4):550--562, 2001.

\bibitem{Gonzalez:06}
G.~Gonzalez.
\newblock {\em Foundations of Oscillator Circuit Design}.
\newblock Artech, 2006.

\bibitem{Gottlieb:97}
I.~Gottlieb.
\newblock {\em Practical Oscillator Handbook}.
\newblock Elsevier, 1997.

\bibitem{HDS:74}
M.~W. Hirsch, R.~L. Devaney, and S.~Smale.
\newblock {\em Differential Equations, Dynamical Systems, and Linear Algebra}.
\newblock Academic press, 1974.

\bibitem{Ijspeert:08}
A.~J. Ijspeert.
\newblock Central pattern generators for locomotion control in animals and robots: a review.
\newblock {\em Neural Networks}, 21(4):642--653, 2008.

\bibitem{Iwasaki:08}
T.~Iwasaki.
\newblock Multivariable harmonic balance for central pattern generators.
\newblock {\em Automatica}, 44(12):3061--3069, 2008.

\bibitem{JF:25}
O.~{Juarez-Alvarez} and A.~Franci.
\newblock Collective rhythm design in coupled mixed-feedback systems through dominance and bifurcations.
\newblock {\em IEEE Transactions on Control of Network Systems}, pages 1--12, 2025.
\newblock (early access).

\bibitem{KAS:99}
H.~Kimura, S.~Akiyama, and K.~Sakurama.
\newblock Realization of dynamic walking and running of the quadruped using neural oscillator.
\newblock {\em Autonomous Robots}, 7(3):247--258, 1999.

\bibitem{KB:50}
N.~M. Krylov and N.~N. Bogoliubov.
\newblock {\em Introduction to Non-Linear Mechanics}.
\newblock Princeton University Press, 1950.

\bibitem{MB:75}
A.~Mees and A.~Bergen.
\newblock Describing functions revisited.
\newblock {\em IEEE Transactions on Automatic Control}, 20(4):473--478, 1975.

\bibitem{MFS:18}
F.~A. {Miranda-Villatoro}, F.~Forni, and R.~Sepulchre.
\newblock Analysis of {L}ur’e dominant systems in the frequency domain.
\newblock {\em Automatica}, 98:76--85, 2018.

\bibitem{SKW:23}
Y.~Sato, Y.~Kawano, and N.~Wada.
\newblock Parametrization of linear controllers for $p$-dominance.
\newblock {\em IEEE Control Systems Letters}, 7:1879--1884, 2023.

\bibitem{VG:02}
S.~Varigonda and T.~T. Georgiou.
\newblock Dynamics of relay relaxation oscillators.
\newblock {\em IEEE Transactions on Automatic Control}, 46(1):65--77, 2002.

\end{thebibliography}

\end{document}